\documentclass[preprint,showpacs,amsmath,prb]{revtex4}
\usepackage{graphicx, latexsym, amssymb}
\newcommand{\cc}[0]{\Large$\circ$\hspace{0.7mm}\normalsize}
\newcommand{\bc}[0]{\Large$\bullet$\hspace{0.7mm}\normalsize}
\newcommand{\tr}[0]{$\triangle$\hspace{0.7mm}}
\newcommand{\bt}[0]{\large$\blacktriangle$\hspace{0.7mm}\normalsize}
\newcommand{\sq}[0]{$\Box$\hspace{0.7mm}}
\newcommand{\bs}[0]{\small$\blacksquare$\hspace{1.5mm}\normalsize}
\newcommand{\di}[0]{$\Diamond$\hspace{0.7mm}}
\newcommand{\pls}[0]{$+$\hspace{0.7mm}}
\newcommand{\crs}[0]{$\times$\hspace{0.7mm}}
\begin{document}
\title{Dimerization-induced enhancement of the spin gap in the quarter-filled two-leg 
rectangular ladder}
\author{Y. Yan} 
\author{S. Mazumdar}
\affiliation{Department of Physics, University of Arizona, Tucson, AZ 85721}
\author{S. Ramasesha}
\affiliation{Solid State and Structural Chemistry Unit, 
Indian Institute of Science, Bangalore 560 012, India}
\date{\today}
\begin{abstract}
We report density-matrix renormalization group calculations of spin gaps in 
the quarter-filled correlated
two-leg rectangular ladder with bond-dimerization along the legs of the ladder. 
In the small rung-coupling region, dimerization along the leg bonds
can lead to large enhancement of the spin gap.
Electron-electron interactions further enhance the spin gap, which is nonzero for 
all values of the rung electron
hopping and for arbitrarily small bond-dimerization.
Very large spin gaps, as are found experimentally in
quarter-filled band organic charge-transfer solids with coupled
pairs of quasi-one-dimensional stacks, however, occur within the model 
only for large dimerization and rung electron hopping that 
are nearly equal to the
hopping along the legs. Coexistence of charge order and spin gap is also possible
within the model for not too large intersite Coulomb interaction. 
\end{abstract}
\maketitle

\section{Introduction}
Coupled pairs of one-dimensional (1D) chains, referred to as two-leg ladders, have been 
widely investigated within the spin 1/2 antiferromagnetic Heisenberg Hamiltonian
because they exhibit behavior different from both the isolated 1D chain as well as the 
two-dimensional (2D) lattice. 
Ignoring electron occupations, two different kinds of couplings between the 
1D chains, as shown in Figs.~1(a) and (b), have been of interest.
We refer to the system
of Fig.~1(a) as a rectangular ladder and that of Fig.~1(b) as a zigzag ladder. 
Within the spin 1/2 Heisenberg Hamiltonian for the rectangular ladder,
the energy gap between
the lowest spin triplet (S=1) excitation and the spin singlet (S=0) ground state,
hereafter the spin gap, is nonzero for  arbitrarily small spin exchange
along the rung bonds. \cite{Dagotto96,Barnes93,White94,Gopalan94} 
The zigzag
ladder of Fig.~1(b) can also be considered as a 1D chain with the interstack
(intrastack) couplings corresponding to the nearest neighbor (next nearest neighbor)
couplings in the 1D chain. 
The spin gap in the antiferromagnetic zigzag ladder is nonzero only for the 
intrastack spin exchange
above a critical value \cite{Tonegawa87,Bursill95,Chitra95,White96}.
Nonzero spin gap is also found in the 1/2-filled band rectangular ladder within the Hubbard
Hamiltonian with finite on-site electron-electron (e-e) interaction.
\cite{White94,Noack96} 
Spin gaps and superconducting pair-pair correlation functions in
weakly doped rectangular ladders have been investigated within 
Hubbard and t-J models \cite{White94}. 


The strong interest in the theory of undoped and weakly doped 1/2-filled band ladders
owes its origin to possible connections with theories of high temperature
superconductivity in the cuprates \cite{Dagotto99}. 
Whether or not significant spin gap  
occurs in ladder structures
at other commensurate bandfillings is also an interesting question.
The particular commensurate bandfilling
that has been investigated so far is 1/4.
Exact diagonalization studies of
finite 1/4-filled rectangular ladders with
e-e and Holstein electron-phonon (e-p) interactions
have demonstrated the coexistence of charge-order and
spin-Peierls states \cite{Riera99}. The magnitude of the spin gap in the thermodynamic
limit due to such a transition was not calculated. 
Density matrix renormalization group (DMRG) calculations for the rectangular
ladder within
an extended Hubbard Hamiltonian with repulsive on-site interaction $U$ and
nearest neighbor interaction $V$ have shown that the checkerboard charge-ordering of Fig.~1(c)
occurs only for $V$ larger than a critical value $V_c(U)$ \cite{Vojta01}.
The calculated spin gaps in the thermodynamic limit, obtained from extrapolations,
are nonzero for $t_{\perp} < t$, where $t_{\perp}$ and $t$ are the one-electron
hopping parameters along the ladder rungs and legs, respectively,
in agreement with weak coupling renormalization group calculations \cite{Balents96}.
The magnitudes of the extrapolated spin gaps are, however, tiny. 
Furthermore, the uncertainties in the
extrapolated spin gaps are rather large in this parameter regime due to finite-size effects, and 
are comparable to the magnitudes of the spin gaps.
 \cite{Noack96,Vojta01} For example, for $U$ = 8, $V$ = 0 and hopping parameter $t_{\perp}$ = 0.7 (all in
units of $t$) the spin gap is $\sim$ 0.03 $\pm$ 0.01, while for the same $U$ and hopping parameter $t_{\perp}$ but
$V$ = 2, the spin gap is $\sim$ 0.01 $\pm$ 0.005.
The spin gap also goes to 0 for hopping parameter $t_{\perp} \geq t$,
and consequently,  
significant spin gap is absent in the regular rectangular 1/4-filled band ladder 
for repulsive $U > V$. Significant spin gap does occur in the zigzag 1/4-filled band
electron ladder with both e-e and e-p interactions. For $t_d > 0.5858t_s$ 
the ground state is a Bond-Charge-Density wave (BCDW) (see Fig.~1(d)) with
coupled lattice distortion, charge order and very large spin gap. 
\cite{Clay05} 

Experimentally, large spin gaps have been observed at low temperatures in several 
recently discovered
organic charge-transfer solids with strong pairwise couplings between 
quasi-1D stacks of organic molecules, and weak interpair couplings
\cite{Ribas05,Ribera99,Wesolowski03,Nakamura02}. 
In all cases, there occurs first a 
metal-insulator transition at high temperature that is accompanied by leg
bond-dimerization, following which
there occurs an insulator-insulator transition which is
accompanied by the opening of a spin gap. The temperatures T$_{SG}$ 
at which the spin gap transitions occur,
$\sim$ 70 K in the charge-transfer solid (DT-TTF)$_2$[Au(mnt)$_2$] \cite{Ribas05}, and $\sim$ 170 K in 
(BDTFP)$_2$PF$_6$(PhCl)$_{0.5}$ \cite{Nakamura02}, are unusually high compared to the
spin-Peierls transition temperatures T$_{SP} \sim$ 15--20 K in the 
1D 1/4-filled band charge-transfer solids \cite{Pouget97,Visser83}. 

The very large T$_{SG}$ in the the coupled-stack systems suggest that the mechanism by
which the gap opens is probably unrelated to the intrachain spin-Peierls transition, which
in the 1/4-filled band is tetramerization, or dimerization of the dimerization. \cite{Visser83} 
Based on the experimentally observed high temperature dimerization, the ladder model
of Fig.~1(d) has been suggested for these systems 
\cite{Ribas05,Ribera99,Wesolowski03,Nakamura02}.
Within this picture, each
dimer unit cell acts as a single site in an {\it effective}
spin 1/2 rectangular ladder with significant spin gap (see below).
The mapping from the dimerized 1/4-filled band electron model to the spin model, 
or the occurrence
of spin gap in the original 1/4-filled band model have, however, not been demonstrated.

Yet another
1/4-filled band ladder system in which relatively large spin gap (34 K) has been observed is
$\alpha'$-NaV$_2$O$_5$, which consists of V-O ladders coupled through weak
V-V bonds, with the V-O layers separated by layers of Na$^+$ ions \cite{Isobe96}. 
Charge order involving the V sites
and spin gap occur at the same temperature in this system.\cite{Isobe96}
Theoretical description 
based on the 
V-only 1/4-filled d-band extended Hubbard model predict 
the checkerboard charge-order pattern of Fig.~1(c). 
for the V-ladders.
\cite{Seo98,Mostovoy00}
The calculated spin gap within the checkerboard charge-order phase for hopping parameter $t_{\perp}>t$, as is
appropriate \cite{Seo98,Mostovoy00} for
$\alpha$-NaV$_2$O$_5$, is however zero \cite{Vojta01}. 
The spin gap here most likely originates from the coupling between the 1/4-filled ladders
\cite{Mostovoy00,Smolinski98,Bernert02,Edegger05}, which can lead to effective dimerization
along the legs of the individual ladders with the checkerboard charge order (see below).

In view of the interest in theoretical models for coupled 1/4-filled band two-chain
systems that give large spin gaps, we have performed DMRG calculations 
for the 1/4-filled band rectangular ladder with leg-bond dimerization 
within the extended Hubbard model. The purpose of our work is not necessarily to
explain the observed experimental behavior of the organic charge-transfer solids or 
$\alpha'$-NaV$_2$O$_5$, but to determine the plausibility of the application of the
models to these systems. The minimum requirement of the applicability of the model
is that significant spin gap occurs within it for realistic rung hopping parameter $t_{\perp}$.
We have considered
two distinct parameter regimes, (i)  $V < V_C(U)$, corresponding to 
homogenous charge distribution on the sites, and (ii) $V > V_c(U)$, which
gives the checkerboard charge order in the infinite ladder. \cite{Vojta01}
In (i), we have considered mostly hopping parameter $t_{\perp} \leq t$, as the
results for larger $t_{\perp}$ are easily anticipated from these calculations. In (ii) we have
considered hopping parameter $t_{\perp} \geq t$ because of the possible applicability of to $\alpha'$-NaV$_2$O$_5$,
in which the rung hopping is larger than the hopping along the leg bonds.
In both cases we find spin gaps that are nearly an order of magnitude
larger than those in the undimerized ladders. 

The paper is organized as follows. In section II we present the theoretical model and methodology.
In section III we show the numerical results for the two cases, nearest neighbor electron-electron interaction $V < V_C(U)$ and  $V > V_c(U)$.
Finally, in section IV we present our conclusions and discuss possible applications of the present work to the 1/4-filled
band paired-stack charge-transfer solids and $\alpha'$-NaV$_2$O$_5$.

\section{Theoretical model and methodology}

We have performed  
DMRG calculations
for the dimerized rectangular ladder of Fig.~1(d), within the Hamiltonian,

\begin{eqnarray}
H =&&U\sum_{i} n_{i,\lambda,\uparrow}n_{i,\lambda,\downarrow} + V\sum_{i,\lambda=-1,+1} n_{i,\lambda}n_{i+1,\lambda} \nonumber \\
   &&+ V_{\perp}\sum_in_{i,1}n_{i,-1} \nonumber \\
	 &&+ t\sum_{i,\sigma,\lambda=-1,+1}(1\pm\delta)(c^\dagger_{i,\lambda,\sigma}c_{i+1,\lambda,\sigma} + h.c.) \nonumber \\
	 &&  + t_{\perp}\sum_{i,\sigma}(c^\dagger_{i,1,\sigma}c_{i,-1,\sigma} + h.c.)
	 \label{extHub}
\end{eqnarray}

Here 
$c^\dagger_{i,\lambda,\sigma}$ creates an electron with spin 
$\sigma$ = $\uparrow, \downarrow$
on site $i$ in ladder leg $\lambda= \pm1$, 
$n_{i,\lambda,\sigma} = c^\dagger_{i,\lambda,\sigma}c_{i,\lambda,\sigma}$, 
and $\delta$ is the bond-alternation
parameter along the legs. The parameters $U, V, V_{\perp}, t, t_{\perp}$ have their
usual meanings. In the following we present all
quantities with dimensions of energy 
in units of hopping parameter $t$. We are
interested in the parameter space  
$U > V,V_{\perp} \geq$ 0, and only in possible spin gap, as the conditions for charge gap 
and charge order 
have been discussed extensively in previous work \cite{Noack96,Vojta01}. 
For simplicity, we present results for $V = V_{\perp}$ and refer to both
interactions as $V$. 
The spin gap $\Delta_S$
is defined as usual as 
E(1) -- E(0), 
where E(S) is the lowest energy in the state with 
total spin S. 
Note that for the 1/4-filled band spin gap $\Delta_S=0$ for hopping parameter $t_{\perp}=0$
even with bond-alternation parameter $\delta \neq 0$. 
Although we are mostly interested in determining whether the spin gap is nonzero 
for arbitrarily small hopping parameter $t_{\perp}$ and bond-alternation parameter $\delta$, our calculations span both
$t_{\perp} < 1$ and $t_{\perp} > 1$. 

Before discussing our numerical results we present here brief discussions of the
the physical mechanism behind the opening up of the spin gap in the dimerized ladder.
Consider first the $U,V,V_{\perp}=0$ band limit of Eq.~\ref{extHub}.
The electronic structure 
in this limit and for
bond-alternation parameter $\delta$ = 0 is given by
noninteracting bonding and antibonding bands split by
2$t_{\perp}$. For $t_{\perp}<1$ and bandfilling of 1/4, electrons occupy both bands and
there are four Fermi points. 
For $t_{\perp}>1$, the lower 
band is 1/2-filled and the upper band empty.
$U>0$ introduces a charge gap now, but the spin gap continues to be zero. 
Nonzero leg bond dimerization $\delta \neq$ 0 
in the limit $U,V,V_{\perp}=0$  
opens a gap at 
the wavevector $k = \pi/2a$ (where $a$ = lattice spacing in the ladder leg direction).
For small $t_{\perp}$ and bond-alternation parameter $\delta$, 
both bands are again occupied.
For $t_{\perp}\sim1$ now there appears a charge and spin 
gap 2$(t_{\perp} - 1 + \delta)$ in the spectrum. For $t_{\perp} >> 1$, the energy gap
at the Fermi surface in the bonding band is 
4$\delta$. These known results suggest that there exists
a range of $t_{\perp}$ and $\delta$ where the spin gap can conceivably be significant for nonzero
$U,V,V_{\perp}$.

The above physical picture does not explain the logic behind the proposed
relationship between the dimerized 1/4-filled
band model of Eq.~(1) and the spin or 1/2-filled band ladders. 
\cite{Ribas05,Ribera99,Wesolowski03,Nakamura02} This can be understood starting from
a different limit, large bond-alternation parameter $\delta \sim 1$. 
In this limit each dimer unit forms a molecule with bonding
and antibonding molecular orbitals (MOs) at energy $\pm 2t$. The single electron within each dimer 
molecule occupies the bonding MO, which forms an effective single site, as the
unoccupied antibonding MO is too high in energy and does not play significant
role in the low energy behavior of the system. The singly
occupied bonding MOs, 
coupled through the weak intersite leg bonds and the rung bonds then constitute
the effective sites of a 1/2-filled band ladder, for which we expect nonzero spin gap for 
arbitrarily small
hopping parameter $t_{\perp}$. 
Whether or not the spin gap is nonvanishing also for small $\delta$ can be found only 
numerically. We will show that the spin gaps that we obtain numerically are considerably more
enhanced compared to those obtained in the $U=V=0$ limit, viz., 2$(t_{\perp} - 1 + \delta)$
for $t_{\perp}\sim1$ and 4$\delta$ for  $t_{\perp} >> 1$. We will also show that the
U-dependence of the spin gaps for the dimerized 1/4-filled band ladder are very similar to those
of the 1/2-filled band ladder.

We calculated the spin gap for rectangular ladders up to $2\times64$ sites
using the infinite-system DMRG algorithm with open boundary condition. Instead of a single
site, a single rung is added to the building block each time. 
The number of states we keep
is m = 300 for the S=0 state and 600 for the S=1 state. We have
confirmed that the  
accuracy of the singlet state 
is comparable to that of the triplet with these m. We 
checked some of our results against published results, 
\cite{Noack96} and found that the errors in our calculations
are comparable, estimated to be less than a few percent. 

In principle, our calculations should
be done with the finite-system DMRG algorithm, which is known to be more
accurate than the infinite-system DMRG algorithm. The former procedure, however,
requires considerably larger amount of time for each different ladder size and parameter
set. The particular problem we consider requires that calculations are done with many different 
$U$, $V$, $t_{\perp}$ and bond-alternation parameter $\delta$ (see below), with 
multiple system sizes $L$ for effective $L \to \infty$ extapolations. The enormous 
amount of computational time that would be necessary for performing all the calculations that are
reported below makes the finite-system DMRG algorithm
impractical in the present case. We note, however, that
in nearly all the cases we discuss below the calculated spin gaps are large
and easy to detect. We
also note that the primary purpose of our work is to determine whether or not significant
spin gaps are obtained within the dimerized ladder, and the precision required is not the same
as within models where the gaps are tiny. \cite{Vojta01} We have
calculated a few of the spin gaps for several different system sizes,
for representative parameters that cover both $V < V_c(U)$ and
$V > V_c(U)$ 
within the finite-system DMRG algorithm using the same number of states as the infinite-system algorithm, and we compare these values with those obtained
with the infinite-system algorithm in Table 1. As seen from the Table, the accuracies of the
infinite-system algorithm are acceptable for our purpose, in both parameter regimes.

\section{Numerical results}

\subsection{Case 1. $V < V_c(U)$}


In Fig. 2(a) we show our calculated spin gap $\Delta_S$
for 1/4-filled 2$\times$L ladders, for the representative case of $U=8$, $V=0$, $t_{\perp}=1$ 
and $0 \leq \delta \leq 0.05$.
For this moderately large $t_{\perp}$, the
accuracy of the numerical results is very high and reliable results are obtained for the
smallest bond-alternation parameter $\delta$.
The kink in the spin gap $\Delta_S$ plot for
bond-alternation parameter $\delta$ = 0.05 is real, - similar behavior has been seen previously for the undimerized
1/4-filled band ladder
(see Fig.~3 in reference \onlinecite{Vojta01}).
The dashed lines indicate the $L \to \infty$ extrapolations of the spin gaps, 
obtained by fitting the calculated spin gap $\Delta_S$ against a 
polynomial in $1/L$ 
up to $1/L^2$. 
For bond-alternation parameter $\delta$ = 0.02 and 0.05, we have also performed
the extrapolations by retaining terms up to $1/L^3$. The two different extrapolations in each of
these cases give the uncertainties in the spin gaps, which are 0.121$\pm$0.008 and 0.244$\pm$0.015 respectively. For all other cases
fittings with $1/L$ alone gave extrapolations that are indistinguishable from the plots
shown on the scale of the figure.
The $L \to \infty$ extrapolated spin gap for bond-alternation parameter $\delta$ = 0 is zero in Fig.~2(a), 
in agreement with previous work.\cite{Vojta01} 
For bond-alternation parameter $\delta$ as small as 0.01, the extrapolated 
spin gap $\Delta_S$ is nonzero for $t_{\perp}=1$. 
Furthermore, for this $t_{\perp}$ the spin gap at $U$ = 0
is 2$\delta$. As seen in the Fig., for all bond-alternation parameter $\delta$, the extrapolated spin gap $\Delta_S$ for
$U$ = 8 is larger. We therefore conclude that the spin gap is enhanced by the on-site
e-e interaction, and is hence nonzero for the smallest bond-alternation parameter $\delta$ at $t_{\perp}$ = 1.
The inset in Fig.~2(a) shows the weakly sublinear behavior of spin gap $\Delta_S$ 
against bond-alternation parameter $\delta$.

Unfortunately, as the system size approaches the thermodynamic limit, the spin gap $\Delta_S$ decreases at smaller
$t_{\perp}$ and the numerical accuracy of the DMRG results at large $L$ are reduced. 
This has also been seen in previous calculations. \cite{Noack96,Vojta01}
Even for
the relatively simpler case of the 1/2-filled Hubbard ladder
precise calculations of the spin gap are difficult for $t_{\perp} < 1$.
\cite{Noack96} Our calculations of the spin gap for the smaller  
$t_{\perp}$ = 0.8 in Fig.~2(b) are therefore for moderate bond-alternation parameter $\delta \geq 0.05$.
Considering that at $U$ = 0 the spin gap $\Delta_S$ = 2($t_{\perp}-1+\delta$) are zero for 
bond-alternation parameter $\delta \leq 0.2$,
we see that the extrapolated
spin gaps are once again strongly enhanced by $U$.
Taken together
with the weak-coupling result that the spin gap is nonzero (albeit small)
for all $t_{\perp} < 1$ at $\delta=0$,\cite{Balents96} the
results of Fig.~2(a) and (b) suggest that for nonzero  $\delta$ and $U$ the spin gap is nonzero for 
all $t_{\perp}$. 
The main difference between $\delta=0$ and $\delta \neq$ 0 is 
that the spin gap in the latter case is
largest for $t_{\perp} \geq$ 1, exactly the region where the spin gap is zero for
$\delta$ = 0. The appearance of nonzero spin gap for the smallest bond-alternation parameter $\delta$ 
indicates that there is indeed a qualitative similarity between the
the 1/4-filled band dimerized rectangular ladder and an effective
spin ladder.

In Fig.~2(c) we have shown the effect of nonzero $V$ on the spin gap for $U = 8$ and
$t_{\perp}$ = 0.8 and bond-alternation $\delta$ = 0.05. $V$ enhances the spin gap very strongly, such that 
even for $t_{\perp}$ = 0.8 the spin gap is now quite large. The enhancement of the spin gap
due to $V$ can be understood physically, given the $\delta$-dependence of spin gap $\Delta_S$.
Defining the bond-orders along the ladder legs in the usual manner,
$B_{i,\lambda} = \langle c^\dagger_{i,\lambda,\sigma}c_{i+1,\lambda,\sigma} 
+ h.c.\rangle$, we note that for nonzero bond-alternation $\delta$ the absolute value of the
difference between consecutive
bond orders, $\Delta B = |B_{i,\lambda} - B_{i+1,\lambda}| > 0$. $\Delta B$ is then
a {\it wavefunction measure of leg bond-dimerization},\cite{Clay05} and 
for fixed bond-alternation $\delta$ we expect the spin gap to
increase with $\Delta B$. We have calculated the exact $\Delta B(V)$ at 
fixed $U=8$ for finite 1/4-filled periodic 1D rings and ladders with $t_{\perp}$ = 0.0 and
0.8 with 12 and 16 sites.
In Fig.~2(d) we have plotted $\Delta B(V)/\Delta B(0)$ for $U=8$ against $V$. 
As a consequence of the well known 4n vs. 4n+2-electron effect, we expect the curves
for the infinite system to be bounded by the curves for the 12- and 16-site periodic systems
for both $t_{\perp}$. 
The enhancement of $\Delta B(V)$ with $V$ indicates enhanced effective leg bond-dimerization
with $V$. 
Thus $V$ increases the spin gap because it increases the dimerized nature of the ground state
wavefunction.
For $U=8$, the checkerboard charge order of Fig.~1(c) is obtained \cite{Vojta01} for
$V > V_c(U) \simeq 2.6$ at
bond-alternation parameter $\delta=0$. \cite{Vojta01}.
Our calculations in Fig.~2(c) and (d) are for $V < V_c(U)$. We discuss $V > V_c(U)$
separately.

In order to get more complete understanding of the behavior of the spin gap within Eq.~\ref{extHub}
we have calculated the $L \to \infty$ extrapolated spin gaps for several sets of $U$, $V$,
$t_{\perp}$ and $\delta$. The parameter region of large $t_{\perp} > 1$ and large 
$\delta > 0.2$ will clearly yield large spin gaps and is hence uninteresting. We are interested
primarily in the $t_{\perp}<1$ region, thought to be applicable to the organic
charge-transfer solids.\cite{Ribas05,Ribera99,Wesolowski03,Nakamura02} Also in the context of the charge-transfer solids, 
$\delta \leq 0.1$. \cite{Clay03}
In Fig.~3(a) we have plotted the extrapolated spin gaps $\Delta_S$ versus $t_{\perp}$ for several
different $U$, but $V$ = 0, for bond-alternation parameter $\delta$ = 0.05. 
Because of the large uncertainties in the
calculated $\Delta_S$ for small $t_{\perp}$
our results
in Fig.~3(a) are limited to $t_{\perp} \geq 0.6$. The results for the same $U$ but larger
bond-alternation $\delta=0.1$ are shown in Fig.~3(b). Fig.~3(c) shows the spin gap behavior for
fixed $U$ = 8, with three different 
$V<V_c(U)$. In all cases, and
for all $t_{\perp}$ we find considerable enhancement of the spin gap by e-e interactions,
as is summarized in Fig.~3(d). 
The behavior of the spin gap as a function of $U$ in Fig.~3(d) is qualitatively similar to that
of the uniform 1/2-filled band rectangular
Hubbard ladder (see Fig.~4 in reference \onlinecite{Noack96}), suggesting again
that the spin excitations of the dimerized
1/4-filled rectangular ladder can be obtained from those of the 
rectangular spin-ladder 
(note that increasing atomic $U$ also increases the Coulomb repulsion between two
electrons occupying the same bonding MO of the dimer unit \cite{Chandross99}.
It has been claimed that such mapping can
give rung spin exchange in the effective spin ladder larger than the nearest neighbor
spin exchange
along the legs, even for smaller hoppings along the rungs within the 
quarter-filled band electron ladder.
\cite{Ribas05,Ribera99,Wesolowski03} 
For the dimerized linear chain the transformation from the 
atomic basis to the dimer MO basis \cite{Chandross99}
gives $t/2$ as the effective hopping between dimer MOs along the chain. This remains true
here for the leg bonds.
On the other hand, within the atomic basis of Eq.~(1), there are two rung hoppings between
the dimer units. The overall 
rung hopping between the dimer bonding MOs then
continues to be $t_{\perp}$ after transformation to the dimer MO basis.
Hence for all rung hopping $t_{\perp}\geq$ 0.5 in the 1/4-filled ladder the
rung spin exchange in the effective spin ladder is larger than the leg spin exchange.
We have confirmed this numerically: the $L \to \infty$
spin gap of the 1/4-filled band ladder at a fixed $U$ with any specific $t_{\perp}/t$ is 
reproduced in the 1/2-filled band ladder with same $U$ for much larger $t_{\perp}/t$.

\subsection{Case 2. $V > V_c(U)$}

The calculations reported in the above are for $V<V_c(U)$ and $t_{\perp}\leq 1$. 
For $V>V_c(U)$ and $U,V>>1$, the checkerboard charge order of Fig.~1(c) is obtained in the infinite
ladder, while in finite systems there occur charge-charge correlations corresponding to this charge order.
\cite{Vojta01}
The ``occupied''
sites in Fig.~1(c) behave as charge $e$ particles with spin 1/2. At $\delta=0$,
nonzero spin gap can occur only for small $t_{\perp}$, where a spontaneous Majumdar-Ghosh
dimerization spin gap \cite{Tonegawa87,Bursill95,Chitra95,White96} opens up.\cite{Vojta01} 
There can, however, be no spin gap in this case for $t_{\perp} \geq$ 1.
The situation is
different for nonzero $\delta$, as indicated in our numerical results for large 
$t_{\perp}=1$ in Fig.~4 (a). $V_c(U)$ here for $\delta=0$ is 2.6 \cite{Vojta01}, and
nonzero $\delta$ is expected to slightly enhance $V_c(U)$. For $t_{\perp}=1$,
we find then coexistence of the
checkerboard charge order and spin gap in the region $3\leq V \leq 4$.
With even larger $t_{\perp}=2$
the spin gap varies only weakly with $V$. The energy gap between the bonding and antibonding
dimer MOs are now very large,
and the system behaves here as a simple 1D dimerized Heisenberg chain.

The width of the region over which there occurs coexistence of charge order and spin gap clearly depends on
the bond-alternation. Indeed, had we performed our calculations 
self-consistently with e-p couplings that
modulated the hopping integrals along the leg bonds, as opposed to with rigid bond-alternation $\delta$,
the amplitude of the bond-dimerization would have decreased with $V$ in the $V>V_c(U)$ 
region and the coexistence region would have been narrower. This is known from 
calculations within the 1D 1/4-filled band e-p coupled extended Hubbard model where the 
bond dimerization
decreases with $V$ in the region $V > V_c(U)$ \cite{Clay03}. The same physics can also
be captured from calculations of the bond orders B(V). In Fig.~4(b) we show plots of
$\Delta B(V)/\Delta B(0)$ for the parameters of Fig.~4(a), obtained from exact diagonalizations
of 12 and 16-site periodic ladders with 6 and 8 electrons, respectively. As in Fig.~2(d), the
plots for the infinite system are expected to lie in between the curves for the 12 and
16-site ladders for each $t_{\perp}$. Thus for the infinite ladder, the bond order difference
is expected to first increase with $V$ at $t_{\perp}$ = 1 up to $V \sim 2$, following which
it is expected to decrease in the $2 < V < 4$ region. Similarly, for $t_{\perp}$ = 2 the
bond order difference is expected to be nearly flat in this region of $V$, with the decrease 
occurring at even larger $V$.
In both cases, the qualitative behavior of the bond order
differences and the spin gaps are the same, emphasizing the strong role of the leg bond dimerization.
Similar relationship between the dimerization and the spin gap has also been observed
in a recent calculation on coupled ladders, where the effective dimerization is a consequence
of the charge order on neighboring ladders. \cite{Edegger05}


\section{Discussions and Conclusions}

In summary, the spin gap within the 1/4-filled band rectangular correlated-electron ladder is
nonzero for the smallest bond-alternation, and increases monotonically with 
rung hopping $t_{\perp}$ in the region $0<t_{\perp}<1$.
The qualitative similarity between the dimerized 1/4-field band ladder and the spin ladder 
is easily
understood in the large bond-alternation region, and persists at smaller bond-alternation, as is
seen from the enhancement of the spin gap by dimerization, and the enhancement of the spin gap by both
$U$ and $V$ for $V<V_c(U)$ at fixed dimerization.
For $V>V_c(U)$, there can be coexistence of charge order and spin gap, especially for 
large $t_{\perp}>1$. 

In the context of the charge-transfer solids, the electronic parameters appropriate for quasi-1D cationic systems
are known to be \cite{Clay03} $t \sim$ 0.1 eV, $U/t \sim 6-8$ and $V \sim 2t < V_c(U)$.  
Assuming similar intrastack parameters for the paired-stack charge-transfer solids, 
the very large T$_{SG}$ in these
indicate spin gap $\Delta_S \geq 0.2$.
Assuming that the bond-dimerization $\delta \sim$ 0.1, as is true in
most quasi-1D charge-transfer solids \cite{Clay03}, we see from Fig.~3 that such large spin gaps are obtained only for 
$t_{\perp} \geq 0.8$, and certainly $t_{\perp} < 0.7$ is impossible. Somewhat larger
spin gap is obtained within the 1/4-filled band zigzag ladder for realistic e-e and e-p
interaction parameters \cite{Clay05}, but the BCDW here appears only for 
$t_d \geq 0.6$. Thus both the electron ladder models would require
substantial interchain hopping to give the observed large spin gaps. The difference between
the two models is that while no charge order is expected within the rectangular ladder
model for the charge-transfer solids, spin gap in the zigzag ladder is accompanied by strong charge order. Experiments that
probe charge disproportionation are required to determine which of the two models apply to
the real materials. 

As regards $\alpha'$-NaV$_2$O$_5$, our numerical results of Fig.~4(a) are 
in agreement to the conclusions of references \onlinecite{Bernert02} and 
\onlinecite{Edegger05}. 
As shown in these papers, because of the coupling between the ladders in $\alpha'$-NaV$_2$O$_5$,
the charge order can drive effective bond dimerization along the ladder legs
that is manifested
in the alternation of intraladder coupling parameters
(there is a subtle difference
between reference \onlinecite{Bernert02}, which models $\alpha'$-NaV$_2$O$_5$ as a system
with charge order on only
alternate ladders and reference \onlinecite{Edegger05} which assumes charge order in every ladder; this
difference is not important in the present context.) The spin gaps in these references are explained within 
coupled dimerized spin chain models. DMRG calculations within the extended Hubbard model 
for two coupled ladders with charge order also find spin gap up on extrapolation \cite{Edegger05}, 
but by necessity the maximum ladder length (20 rungs) is small  here. Our calculations are thus
complementary to previous work, showing that once it is assumed that interladder coupling
gives rise to effective bond dimerization the essential physics is captured within the
electronic Hamiltonian for the single ladder.

\section{Acknowledgments}

S.M. acknowledges useful discussions with R. Torsten Clay and J. Musfeldt. This work
was supported by the NSF-DMR-0406604, NSF-INT-0138051 and DST, India through 
/INT/US(NSF-RP078)/2001.

\clearpage
\newcommand{\lb}[1]{\raisebox{-2mm}[0pt]{#1}}
\newcommand{\rb}[1]{\raisebox{1.5mm}[0pt]{#1}}
\newcommand{\hsp}[0]{\hspace{0.5cm}}
\newcommand{\colL}[0]{\hsp L \hsp}

\begin{table}[htb]
  \caption{Spin gaps obtained by using the infinite-system and finite-system DMRG algorithms. In all cases $U$=8.}

  \begin{tabular}{|c|c|c|c|}
     \hline
        \lb{Parameters}                 & \lb{\colL}& \multicolumn{2}{c|}{spin gap $\Delta_s$}       \\  \cline{3-4}                          
                                        &           & infinite          & finite            \\  \hline   
                                        & 40        & 0.178337          & 0.177095          \\                               
        $V$=0                           & 48        & 0.165737          & 0.164359          \\                               
        $t_\perp$=1, $\delta$=0.02      & 56        & 0.157386          & 0.155845          \\  
                                        & 64        & 0.151652          & 0.149933          \\  
                                        & $\infty$	& 0.121$\pm$0.008	  & 0.111$\pm$0.006		\\ \hline

                                        & 40        & 0.235359          & 0.234463          \\  
        $V$=2                           & 48        & 0.234640          & 0.233727          \\  
        $t_\perp$=0.8, $\delta$=0.05    & 56        & 0.234241          & 0.233316          \\  
                                        & 64        & 0.234000          & 0.233068          \\ 
                                        & $\infty$	& 0.233$\pm$0.001 	& 0.232$\pm$0.001 	\\ \hline

                                        & 40        & 0.166847          & 0.166470          \\  
        $V$=3                           & 48        & 0.162380          & 0.161969          \\  
        $t_\perp$=1, $\delta$=0.05      & 56        & 0.159576          & 0.159126          \\  
                                        & 64        & 0.157700          & 0.157209          \\
                                        & $\infty$	& 0.146$\pm$0.003		& 0.145$\pm$0.003		\\ \hline
    \end{tabular}
\end{table}

\clearpage
\centerline{Figure Captions}
\vskip 1pc
\noindent Figure 1. Quarter-filled band two-leg electron ladders. 
Grey, black and white circles represent 
mean electron occupations
0.5, $>$ 0.5 and
$<$ 0.5, respectively.
(a) Rectangular ladder with hopping integrals $t$ and $t_{\perp}$ along
the leg and the rung, respectively. (b) The BCDW broken symmetry in the
1/4-filled zigzag ladder, with average intrastack and interstack hopping integrals
$t_s$ and $t_d$, respectively.
(c) The checkerboard charge order in the 1/4-filled band
rectangular ladder.
(d) The bond-dimerized 1/4-filled band rectangular ladder.
\vskip 1pc

\noindent Figure 2 (a) and (b) Convergence behavior of spin gaps for the 1/4-filled band
dimerized rectangular ladder for $U$=8 and $V$=0. In (a) the rung hopping parameter $t_{\perp}$=1, the maximum number of rungs $L_{max}$=80;
\cc, \tr, \sq and \di 
correspond to bond-alternation parameter $\delta$=0, 0.01, 0.02 and 0.05, respectively. In the curves corresponding to 
$\delta$=0.02 and 0.05, the upper L $\to$ $\infty$  extrapolation is using polynomial in $\frac{1}{L}$ up to $\frac{1}{L^3}$ and the lower one up to $\frac{1}{L^2}$.

In (b),
$t_{\perp}$=0.8; 
\di, \pls and \crs
correspond to $\delta$=0.05, 0.1 and 0.2, 
respectively. (c) Spin gaps for $U$=8, $t_{\perp}$=0.8, $\delta$=0.05, with different $V$. 
\di, \bt and \bs
correspond to $V$=0, 1 and 2, respectively. (d) The normalized 
absolute 
difference in consecutive bond orders, $\Delta$B(V)/$\Delta$B(0), as a function of $V$ for
$U$ = 8. Here 
\cc and \sq
correspond to periodic rings
with 12 and 16 sites (i.e., $t_{\perp}$=0 in Eq.~\ref{extHub} and $\delta$=0.05) , respectively; the 
\bc and \bs to ladders with 12 and 16 sites ($t_{\perp}$=0.8, $\delta$=0.1) respectively. 
In all cases the number of electrons is half the number of sites.
\vskip 1pc

\noindent Figure 3 (a) DMRG $L \to \infty$ spin gaps versus rung hopping $t_{\perp}$ for different $U$, for
bond-alternation $\delta$=0.05. (b) Same for $\delta$=0.1. 
(c) Same for fixed $U$=8 with different $V$.
(d) Extrapolated spin gaps at $\delta$=0.05 versus 
$U$, for three different $t_{\perp}$. 
\bc and \bs
correspond to
$V$=1 for rung hoppings $t_{\perp}$=0.8 and $t_{\perp}$=1. 
\vskip 1pc
\noindent Figure 4.
(a) $L \to \infty$ spin gaps versus $V$ for $U$=8, $\delta$=0.05 
and two values of $t_{\perp}$. For
$V\geq 3$ the spin gap coexists with the checkerboard charge order. (b) The normalized
absolute difference in consecutive bond orders;
The \cc and \sq correspond to 12 and 16-site ladders,
respectively, at $t_{\perp}$=1. The \bc and \bs correspond to 
12 and 16-site ladders, respectively, at $t_{\perp}$=2.

\clearpage
\begin{figure}[thb]
\centerline{\resizebox{5.4in}{!}{\includegraphics{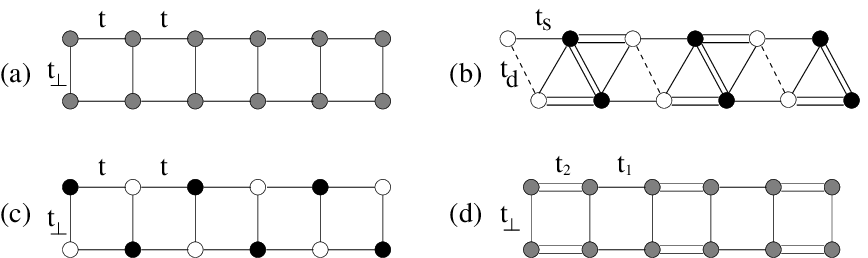}}}
\vskip 0.5 truein
\caption{}
\label{ladders}
\end{figure}

\clearpage
\begin{figure}[thb]
\centerline{\resizebox{5.4in}{!}{\includegraphics{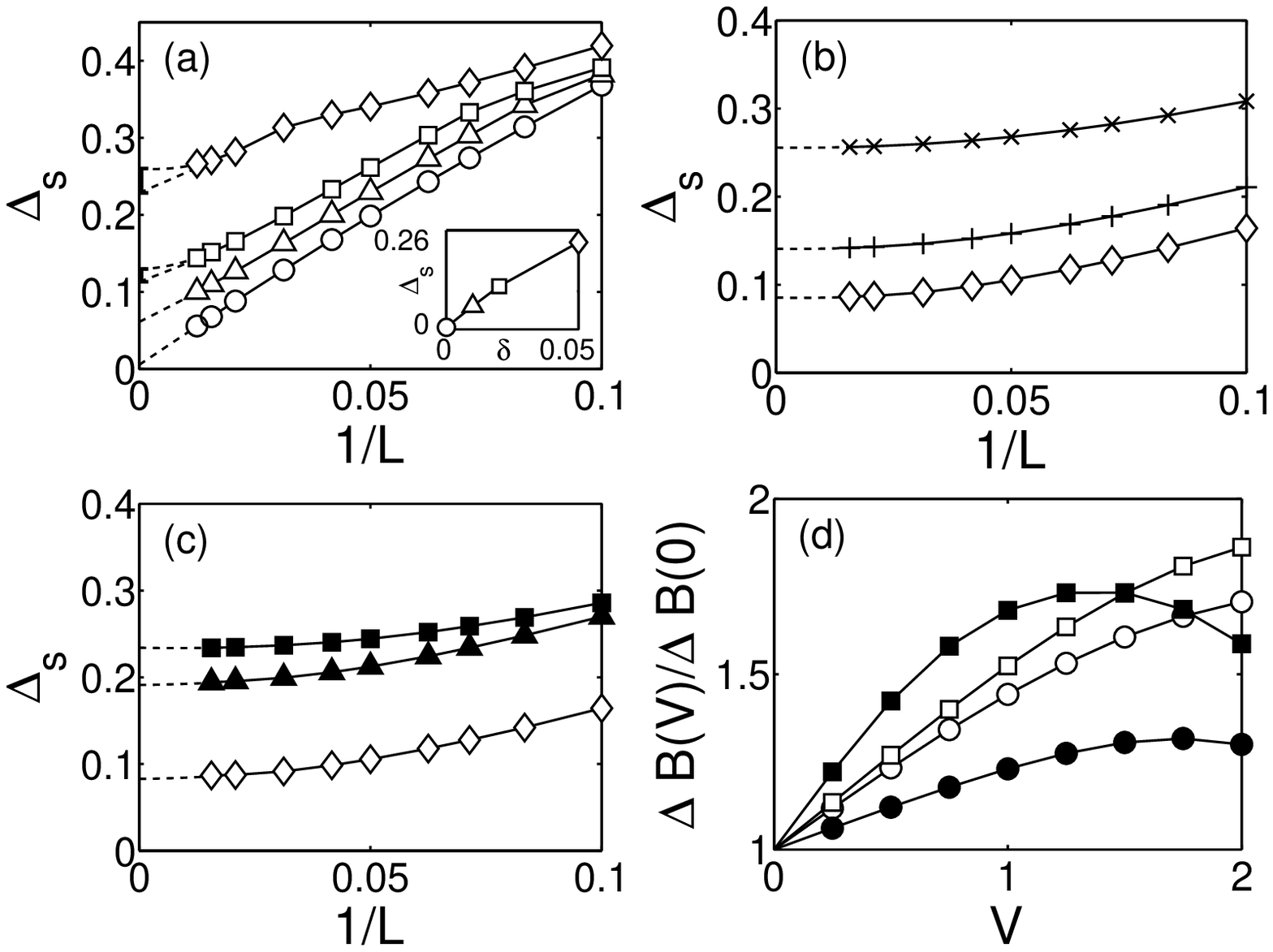}}}
\vskip 0.5 truein
\caption{}
\label{extrapolations}
\end{figure}

\clearpage
\begin{figure}[thb]
\centerline{\resizebox{5.4in}{!}{\includegraphics{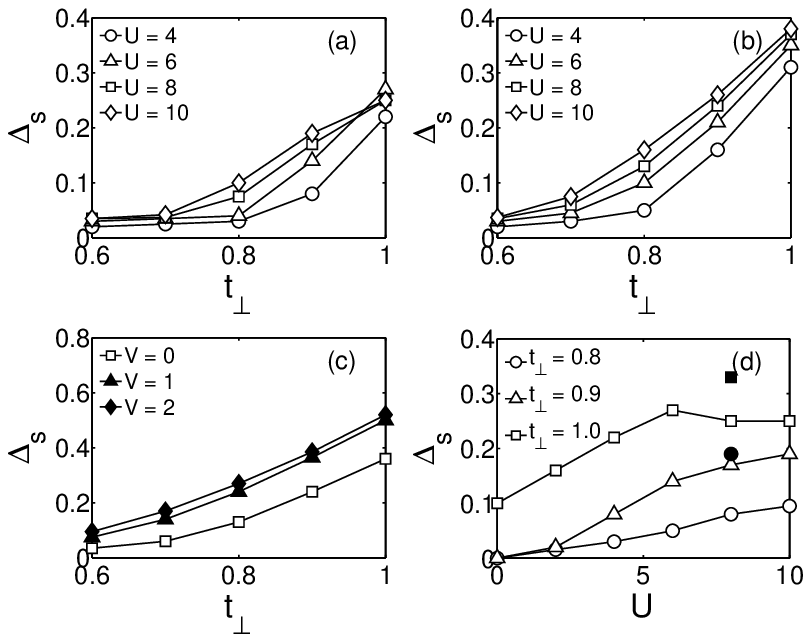}}}
\vskip 0.5 truein
\caption{}
\end{figure}

\clearpage
\begin{figure}
\centerline{\resizebox{5.9in}{!}{\includegraphics{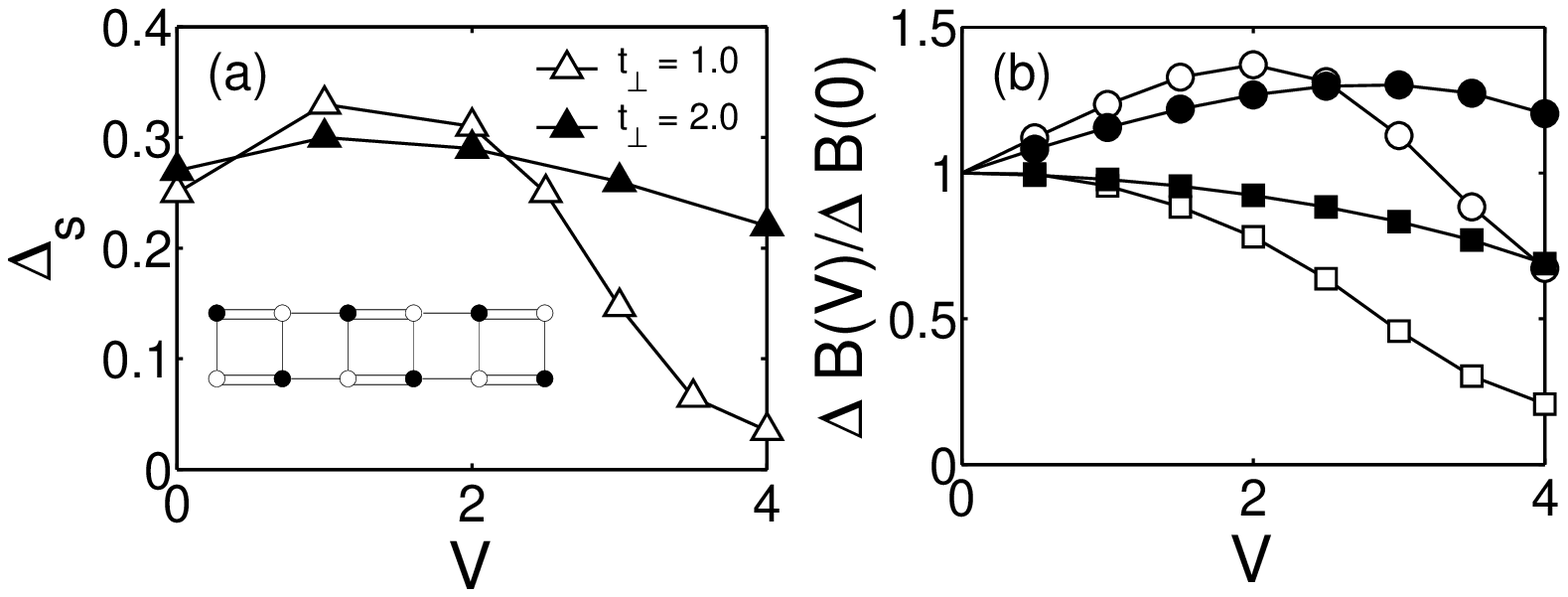}}}
\vskip 0.5 truein
\caption{}

\end{figure}


\begin{thebibliography}{99}
\bibitem{Dagotto96} E. Dagotto and T.M. Rice, Science {\bf 271
}, 618
(1996). 
\bibitem{Barnes93} T. Barnes, E. Dagotto, J. Riera and E. S. Swanson, Phys. Rev. B {\bf 47}, 3196 (1993). 
\bibitem{White94} S.R. White, R.M. Noack
and D.J. Scalapino, Phys. Rev. Lett. {\bf 73}, 886 (1994). 
\bibitem{Gopalan94} S. Gopalan, T.M. Rice
and M. Sigrist, Phys. Rev. B {\bf 49}, 8901 (1994).
\bibitem{Tonegawa87} T. Tonegawa and I. Harada, J. Phys. Soc. Jpn. 
{\bf 56}, 2152 (1987).
\bibitem{Bursill95} R.J. Bursill, G. A. Gehring, D. J. J. Farnell, J. B. Parkinson, T. Xiang and C. Zeng, J. Phys. Condens. Matter {\bf 7}, 8605 (1995).
\bibitem{Chitra95} R. Chitra, Swapan Pati, H. R. Krishnamurthy, Diptiman Sen and S. Ramasesha,  Phys. Rev. B {\bf 52}, 6581 (1995).
\bibitem{White96} S.R. White and I. Affleck, Phys. Rev. B {\bf 54}, 9862 (1996).
\bibitem{Noack96} R.M. Noack, S.R. White and D.J. Scalapino,
Physica C {\bf 270}, 281 (1996).
\bibitem{Dagotto99} E. Dagotto, Reports on Progress in Physics {\bf 62}, 1525 (1999).
The observed superconductivity in the cuprate ladder materials 
(La,Y,Sr,Ca)$_{14}$Cu$_{24}$O$_{41}$ is, however, probably unrelated to theoretical
models of superconductivity in doped spin ladders. See,
T. Vuletic, B. Korin-Hamzic, T. Ivek, S. Tomic, B. Gorshunov, M. Dressel and
J. Akimitsu, Phys. Rep., to be published.
\bibitem{Vojta01} M. Vojta, A. H\"ubsch and R.M. Noack, Phys. Rev. B
{\bf 63}, 045105 (2001).
\bibitem{Riera99} J. Riera and D. Poilblanc, Phys. Rev. B {\bf 59}, 2667 (1999).
\bibitem{Balents96} L. Balents and M..A. Fisher, Phys. Rev. B {\bf 53}, 12133 (1996).
\bibitem{Clay05} R.T. Clay and S. Mazumdar, Phys. Rev. Lett. {\bf 94}, 207206 (2005).
\bibitem{Ribas05} X. Ribas, M. Mas-Torrent, A. Perez-Benitez, J. C. Dias, H. Alves, E. B. Lopes, R. T. Henriques, E. Molins, I. C. Santos, K. Wurst, P. FouryLeylekian, M. Almeida, J. Veciana and C. Rovira, Adv. Funct. Mater. {\bf 15}, 1023 (2005).
\bibitem{Ribera99} E. Ribera, C. Rovira, J. Veciana, J. Tarres, E. Canadell, R. Rousseau, E. Molins, M. Mas, J. Schoeffel, J. Pouget, J. Morgado, R. T. Henriques, M. Almeida and E. Ribera, Chem. Eur. J. {\bf 5}, 2025 (1999).
\bibitem{Wesolowski03} R. Wesolowski, J. T. Haraldsen, J. L. Musfeldt, T. Barnes, M. Mas-Torrent, C. Rovira, R. T. Henriques and M. Almeida, Phys. Rev. B {\bf 68},
134405 (2003).
\bibitem{Nakamura02} T. Nakamura, K. Takahashi1, T. Shirahata2, M. Uruichi, K. Yakushi and T. Mori, J. Phys. Soc. Jpn. {\bf 71},
2022 (2002).
\bibitem{Pouget97} J.P. Pouget and S. Ravy, Synth. Metals {\bf 85},
1523 (1997).
\bibitem{Visser83} R. J. J. Visser, S. Oostra, C. Vettier and J. Voiron., Phys. Rev. B {\bf 28}, 2074 (1983).
\bibitem{Isobe96} M. Isobe and Y. Ueda, J. Phys. Soc. Jpn. {\bf 65}, 1178 (1996).
\bibitem{Seo98} H. Seo and H. Fukuyama, J. Phys. Soc. Jpn. {\bf 67}, 2602 (1998).
\bibitem{Mostovoy00} M.V. Mostovoy and D.I. Khomskii, Solid St. Commun. {\bf 113},
159 (2000).
\bibitem{Smolinski98} H. Smolinski, C. Gros, W. Weber, U. Peuchert, G. Roth,
M. Weiden and C. Geibel, Phys. Rev. Lett. {\bf 80}, 5164 (1998) and references therein.
\bibitem{Bernert02} A. Bernert, P. Thalmeier and P. Fulde, Phys. Rev. B {\bf 66},
165108 (2002) and references therein.
\bibitem{Edegger05} B. Edegger, H.G. Evertz and R.M. Noack, cond-mat/0510325 and 
references therein.
\bibitem{Chandross99} M. Chandross, Y. Shimoi and S. Mazumdar, Phys. Rev. B {\bf 59},
4822 (1999). 
\bibitem{Clay03} R.T. Clay, S. Mazumdar and D.K. Campbell, Phys. Rev. B {\bf 67}, 115121 (2003).
\end{thebibliography}
\end{document}